\documentstyle[11pt,newpasp,twoside,epsf]{article}
\def\edcomment#1{\iffalse\marginpar{\raggedright\sl#1\/}\else\relax\fi}
\reversemarginpar

\begin{document}
\title{TYPE 1 AGN AND THEIR LINK TO ULIRGS}
 \author{Paola Andreani}
\affil{Osservatorio Astronomico di Padova, Padova, Italy\\
Max-Planck Institut f. Extraterrestrische Physik,
Garching, Germany\\
e-mail: andreani@pd.astro.it}

\begin{abstract}
Increasing
observational evidence supports a picture for a close link among
AGN phenomenum, star-formation processes and galaxy formation.
 Since the physical phenomena related to the onset of both
AGN and star formation are very likely characterized by a strong
far infrared (FIR) emission
it becomes mandatory to investigate the FIR
energy domain.  However, because of their faintness in the FIR energy
range, very little is known
about the FIR properties of type 1 AGN, and type 2 AGN,
which are supposed to be
missed in classical searches, still lack definition and samples. 
\hfill\break
Our aim is to investigate
the FIR properties of optically selected
type 1 AGN, compare them to Ultraluminous Infrared Galaxies (ULIRGs) and
derive some general characteristics of the population.
Expectations for future surveys are also presented.
\end{abstract}

\section{Introduction}

For historical reasons and selection criteria, a classical type 1 AGN
is that showing a spectral energy distribution (SED) with most of its energy
in the UV and only 10 - 20 \% in the infrared. From 10\,$\mu$m to
1000 \AA~ the agreement among different composite spectra, built on observations
from different databases, is excellent, while in the FIR
the various studies disagree. The FIR SED differs not only in shape but also the
FIR luminosity differs by up two order of magnitudes. 
This difference is due to inhomogeneities of
AGN samples, covering different luminosity ranges and objects selected
according to different criteria.\hfill\break
However a common result of these studies is that both AGN
and starburst (SB)
 dominated objects have a common FIR-emitting star-formation component
from which is not easy to disentangle pure AGN emission.
Quite a number of observational evidence supports the above
claim: (a)
their 60\,$\mu$m flux, radio and CO emissions are well correlated
and AGN and star-forming galaxies show the same properties;
(b)
their SEDs at $\lambda >$ 60\,$\mu$m have the same
shape as those of star-forming regions;
(c)
the FIR Luminosity Function of AGN coincides with that of
SBs in shape and differ only in amplitude:
$ \Phi^\star_{AGN}/\Phi^\star_{SB}=1/25$;
(d)
the evolution rate of the two classes is comparable and shows
the same behaviour $L(z)\propto (1+z)^{2\div3}$
(e) the FIR component 
represents a distinct component correlating only weakly with
the true AGN emission and very likely related
to a concurrent SB (Lawrence, 2001; Rowan-Robinson, 2001).
\hfill\break
All this supports the idea that all AGN are accompanied by a
SB but one SB out of 25 is accompanied by an AGN.
There are indeed evidences for a coexistence of central AGN
and circumnuclear star formation in a significant fraction
of ULIRGs (Genzel et al., 1998) but AGN are difficult to be observationally
proved mainly because of the high extinction of the nuclear source.
Other distinct direct and indirect observational probes indicate that
an unknown fraction of AGN is absorbed, the so-called type 2,
and the ratio between type 2 and type 1  
may be as high as 10 of which 20 percent in FIR selected objects.
For instance, the comoving mass density  
of Massive Dark Objects, which are probably supermassive
Black Holes in galaxy centers, are a factor of 2 or larger than that explained
by accretion onto optically-luminous AGN. This implies that
much accretion is obscured by dust (Haehnelt, Natarajan and Rees, 1998)
a population which is missing in optically selected samples.
Other evidences
come from X-ray studies and are extensively discussed in this
volume.\hfill\break
Dust enshrouded type 1 AGN exist at high and low redshifts
and constitue a subset of SCUBA galaxies, which in turn
share common properties with local ULIRGs.
Nearly all nearby ULIRGs appear to be late-stage mergers, then
if one out 10 SB has quasar-like activity is then possible to
envisage the evolutionary path starting from a merging system
going through a starburst and ending to a optical-phase QSO
(Sanders et al., 1989).
\hfill\break
\section{Present work}
The aim of the present work is to use FIR observations,
gathered by the Infrared Space Observatory (ISO), of
optical AGN to infer any
evolutionary link from `pure' AGN objects to eventual transition objects
linking AGN to ULIRGs.
\hfill\break
This aim was pursued by comparing FIR and optical properties
in a complete subsample of optically selected bright (15 $\leq$ B $\leq$ 17)
quasars. The sample completeness ensures a better control of observational
biases and provides reliable information on the properties of the underlying
population. The objects studied in this work
can be then fully considered as representative of the entire population.
The optical observations consist of multicolour (U,B,V,R,I)
broad-band photometry
used primarily to selected quasars candidates, which were then
later spectroscopically confirmed.
ISO FIR photometry has been taken at 7,12,25,60,100 and 160\,$\mu$m. 
All the details of these observations and the sample properties
can be found in Andreani et al. (2001).
\section{A colour-colour diagramme for QSOs}
Figure \ref{fig:col_sp}(A) reports $\alpha(60\mu m,{\rm MIR})$ versus
$\alpha({\rm FIR},60\mu m)$. By MIR we indicate the mid-IR bands
7, 12 and 25\,$\mu$m and by FIR the far-IR bands at 100 and 160\,$\mu$m.
 $\alpha$ are the spectral indices, i.e. for a $f_\nu \propto \nu^\alpha$
SED: $ \alpha(\nu_1,\nu_2)=\log (f(\nu_1)/f(\nu_2)) -\log(\nu_1/\nu_2) $
and are therefore simply related to colours.
Our results (see Andreani et., 2001) indicate that $\alpha({\rm FIR},60\mu m)$
is a fair estimate of T$^{-1}_{\rm wd}$, where T$_{\rm wd}$ is
the temperature of a {\tt warm} dust component.\hfill\break
The diagram shows a trend: the relative emission at MIR (shorter) wavelengths
with respect to the 60$\mu$m emission
decreases as T$_{\rm wd}$ increases. A possible interpretation
is the increasing dominance of SB component with respect to AGN
in objects with lower values of T$_{\rm wd}$ and lower values of
the ratio ${\rm F(60\mu m)}/{\rm F({\rm MIR})}$.\hfill\break
The colour-colour diagram was indeed widely used in the past
as a tool to detect and discriminate between different types of activity
in the nuclear and circumnuclear regions of galaxies
(see e.g., Canalizo \& Stockton (2001) and references therein).
Different kind of objects (QSOs/Seyferts, starbursts and ULIRGS) occupy
distinct areas of such a diagram: where FIR fluxes are dominated by dust
reradiation objects locate in the lower right corner (in our
convention at lower T$_{\rm wd}$ and lower $\alpha(60\mu m,{\rm MIR})$)
while those with strong non-thermal emission, such as optically selected QSOs,
are in the upper left corner.
This means that our sample spans a wide range of properties from those
`AGN pure' to those where a concurrent SB dominates.
\section{The type 1 SED}
\begin{figure}
   \plottwo{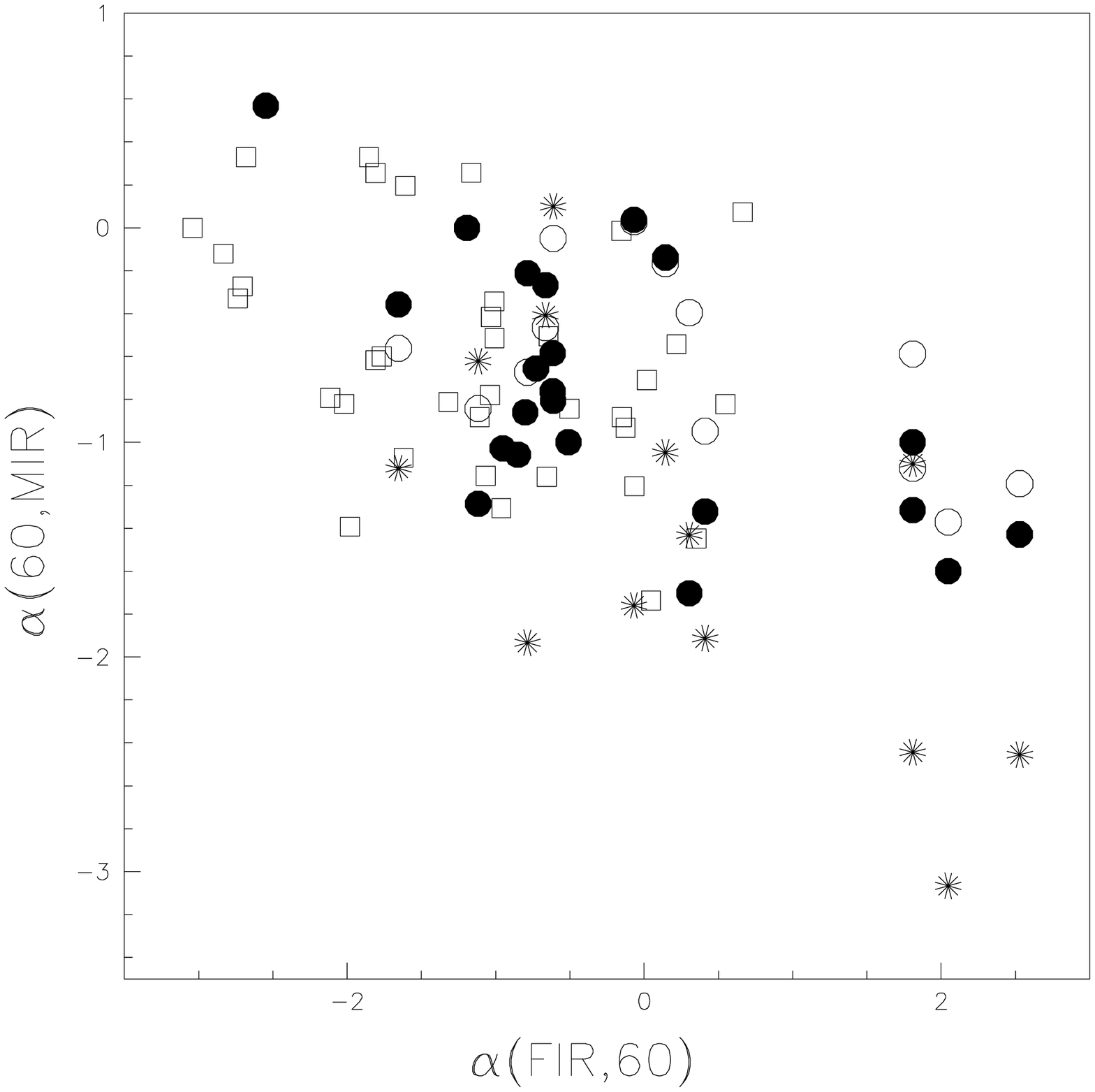}{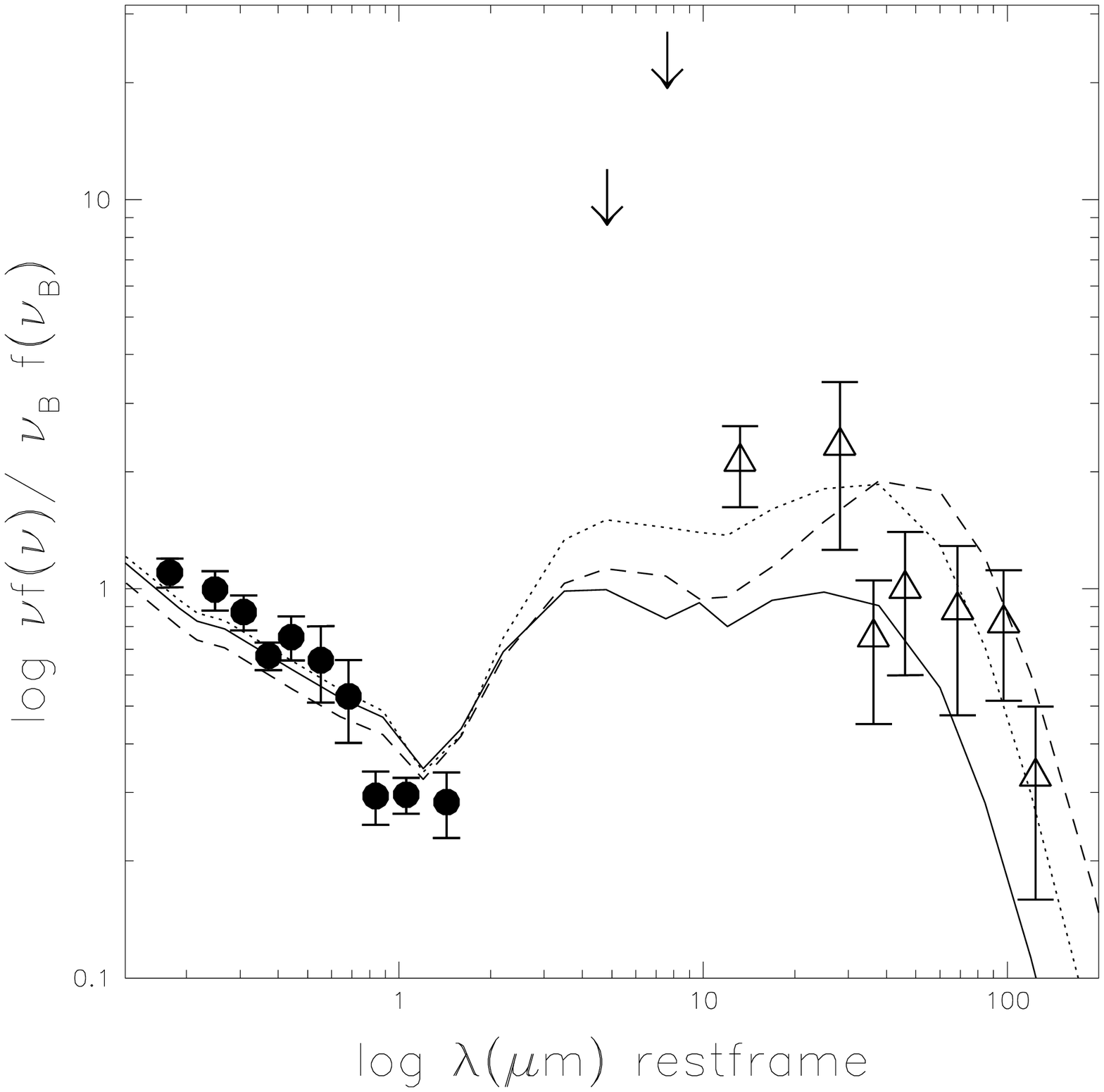}
\caption{Left (A): The ratio between 60$\mu$m flux and MIR fluxes
is shown as a function of the ratio between FIR and 60 $\mu m$ fluxes.
Asterisks correspond to the 25 $\mu$m data, filled squares
to 12\,$\mu$m, open circles to 7\,$\mu$m ones. Open squares correspond
to QSOs at $z \leq 2$ reported in Andreani, Franceschini \& Granato (1999).
Distinct areas can be
identified from AGN dominated (in the upper left corner) to
SB dominated objects (in the lower right corner), according to
the expected dominant emitting mechanism in the object SED from
non-thermal to thermal. Right (B):
The average spectrum in the QSO restframe.
The spectrum is normalized at $\lambda _{\rm blue} =0.44 \,\mu$m.
Filled circles correspond to the optical and near-IR data, open triangles to
the ISO measurements. The lines going through the
data show the predictions of models by Granato and Danese (1994).
The solid line corresponds to a face-on compact configuration
of the torus with absorption at 30$\mu$m of $\tau$=30;
the dotted and dashed lines to a 45$^\circ$
inclined torus with larger dimension and $\tau$=30 and $\tau$=60, respectively.
}
\label{fig:col_sp}
\end{figure}

By making the rather coarse assumption
that optical quasars can thus be modeled as
a homogeneous population, observations at fixed frequencies of
targets at different redshifts can be then used to build the entire
SED for this population.
This analysis exploits the whole spectral information from the optical to
the FIR and helps outlining possible general features of the emitting
mechanisms and the physics of the QSO environment.
\hfill\break
The spectrum is built by dividing the available wavelength range in bins.
In each bin an average of the fluxes at that wavelength is computed.
The errorbars shown are those related to the averages but take into account
also photometric uncertainties.
Figure \ref{fig:col_sp}(B) reports the average spectrum in the QSO restframe.
A well defined IR component peaking at 10-30 $\mu m$ and dropping steeply
above 100\,$\mu$m is evident. Although from the characteristic shape
it is quite straightforward to infer a thermal origin for this component
it can be due either
to a SB emission in the host galaxy and/or to a dusty torus around the
central source. When comparing these data with models (Granato \& Danese,
1994) it is
possible to bound the available parameters on the dust distribution.
It turns out that
the dust distribution around the central source should be compact
because of the sharp cut-off at long wavelengths of the FIR bump.
A larger bump would imply larger temperature distribution
and very likely larger spatial dimension of the dust distribution.
Furthermore, the data are better in agreement with models predicting
the strongest dust self-absorption around 10-30\,$\mu$m.
\section{Expected counts of Type 1 AGN}
\begin{figure}
   \plotone{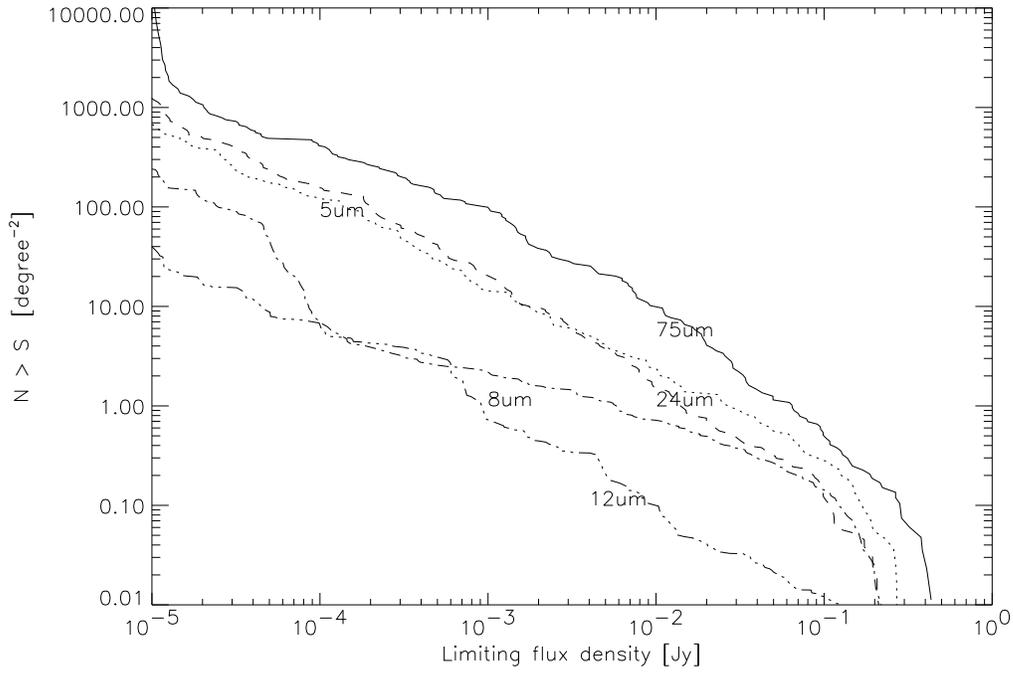}
\caption{Expected number counts of type 1 AGN computed on the basis
of the composite SED of the type 1 AGN of the sample presented in this
work. The different wavelengths correspond to the
bands of IRAC and MIPS instruments on board of SIRT-F and PACS
on board of Herschel. 12$\mu$m counts are only shown for comparison.}
\label{fig:counts}
\end{figure}

Based on these results we show here the predictions of simple models on
the expected numbers of
type 1 QSOs in deep future surveys such those foreseen with
SIRT-F and HERSCHEL Satellites.
\hfill\break
As well known number counts of extragalactic objects represent one of the
fundamental tool to investigate galaxy evolution, since they are related to
the Luminosity Function. The
differential counts, $dN(S)/dS$, can be expressed as a simple integral of the
epoch-dependent luminosity function, $\Psi(L,z)$:
 \begin{equation}
\frac{dN}{dS} = \int ^{z_u} _{z_l} dz \frac{dV}{dz} \frac{dL(S,z)}{dS}
\Psi [L(S,z),z]  ~~~~~~ sources/sr
\end{equation}
\label{eq:difcoun}

\noindent
where $z_u$ and $z_l$ being the effective upper and lower limits of
the redshift distribution and $\frac{dV}{dz}$ the differential
comoving volume.
The integrated counts are easily found by integrating eq.(1)
at different flux limits, $S$.
\hfill\break
As $\Psi(L,z)$ we take the B-band luminosity function that computed
by Boyle et al. (2001) and as its redshift evolution that described
in Bianchi, Cristiani \& Tae-Sum (2001). The latter has
a down-turn at $z=3$ and decays exponentially until $z=10$. The cosmological
parameters used to compute the differential volume are H$_0$=50 km/s/Mpc,
$\Omega=1$, $\Lambda=0$.
We compute k-corrections, i.e. $dL(S,z)/dS$,
according to the composite spectrum of Figure \ref{fig:col_sp}(B)
and plot in Figure \ref{fig:counts}
the expected counts in different observing bands: 5, 8, 24 and 60\,$\mu$m.
i.e. those corresponding to the filter sets of IRAC and MIPS on board
of SIRT-F and PACS on board of Herschel.
It is easily shown that for a deep survey in a sky area of 0.3$\rm deg^2$
at a depth of 0.4 mJy at 24$\mu$m and of 34$\mu$Jy at 8$\mu$m we expect
to see with SIRT-F a few tens and a few hundreds objects, respectively.
Note that the numbers presented here refer to an ideal survey not
affected by observational uncertainties (confusion noise, photometric
errors, ...). More realistic predictions taking into
account also the selection of candidate QSOs
will be the subject of a forthcoming paper.

\begin{table}
\caption{Expected counts of type-1 AGN in future IR surveys}
\begin{tabular}{||cccc|c||}
\hline
\hline

$\lambda$ ($\mu$m) & S$_{\rm min}$($\mu$Jy) & sky area ($deg^2$) &
\# of objects & Instrument\\

\hline

4.5 & 5    & 0.3 & 1000 & IRAC/SIRT-F \\
8.0 & 34   & 0.3 &  350 & IRAC/SIRT-F\\
24  & 370  & 0.3 &   50 & MIPS/SIRT-F\\
75  & 5000 & 1.0 &   22 & PACS/HSO\\

\hline
\hline
\end{tabular}
\end{table}

\end{document}